| | |
|---|---|
| Title | Low-density polyethylene films treated by an atmospheric Ar-$O_2$ post-discharge: functionalization, etching, degradation and partial recovery of the native wettability state |
| Authors | S Abou Rich[1], T Dufour[1], P Leroy[1], L Nittler[2], J J Pireaux[2] and F Reniers[1] |
| Affiliations | [1] Faculté des Sciences, Service de Chimie Analytique et de Chimie des Interfaces, Université Libre de Bruxelles, CP-255, Bld du Triomphe, B-1050 Bruxelles, Belgium<br>[2] Centre de Recherche en Physique de la Matière et du Rayonnement, University of Namur, PMR, 61 rue de Bruxelles, B-5000 Namur, Belgium |
| Ref. | J. Phys. D: Appl. Phys., 2014, Vol. 47, Issue 6, 065203 |
| DOI | http://dx.doi.org/10.1088/0022-3727/47/6/065203 |
| Abstract | To optimize the adhesion of layers presenting strong barrier properties on low-density polyethylene (LDPE) surfaces, we investigated the influence of argon and argon–oxygen atmospheric pressure post-discharges. This study was performed using x-ray photoelectron spectroscopy, atomic force microscopy, optical emission spectroscopy (OES) and dynamic water contact angle (WCA) measurements. After the plasma treatment, a slight increase in the roughness was emphasized, more particularly for the samples treated in a post-discharge supplied in oxygen. Measurements of the surface roughness and of the oxygen surface concentration suggested the competition of two processes playing a role on the surface hydrophilicity and occurring during the post-discharge treatment: the etching and the activation of the surface. The etching rate was estimated to about 2.7 nm s$^{-1}$ and 5.8 nm s$^{-1}$ for Ar and Ar-$O_2$ post-discharges, respectively. The mechanisms underlying this etching were investigated through experiments, in which we discuss the influence of the $O_2$ flow rate and the distance (gap) separating the plasma torch from the LDPE surface located downstream. O atoms and NO molecules (emitting in the UV range) detected by OES seem to be good candidates to explain the etching process. An ageing study is also presented to evidence the stability of the treated surfaces over 60 days. After 60 days of storage, we showed that whatever the $O_2$ flow rate, the treated films registered a loss of their hydrophilic state since their WCA increased towards a common threshold of 80°. This 'hydrophobic recovery' effect was mostly attributed to the reorientation of induced polar chemical groups into the bulk of the material. Indeed, the relative concentrations of the carbonyl and carboxyl groups at the surface decreased with the storage time and seemed to reach a plateau after 30 days. |

# 1. Introduction

Polymers such as polyethylene present excellent physical and chemical bulk properties but exhibit poor surface adhesion features. The deposition of a subsequent layer onto a low-density polyethylene (LDPE) surface can be achieved for specific applications, for instance a layer acting as an oxygen barrier for food packaging. In order to improve the adhesion of a subsequent layer onto this polymer, a pre-treatment of its surface is therefore usually performed. One of the most promising techniques is the plasma activation which—for instance—consists of exposing the sample surface to the post-discharge of a radio frequency (RF) plasma torch. These plasma treatments can be carried out with different gases and vapours like $O_2$, $N_2$, $NH_3$, $H_2O$, $CO_2$, air and noble gases [1–4]. Depending on the experimental conditions, it is considered that a plasma treatment leads to one of the four following effects: cleaning (removing of organic contaminants), etching (removing of polymer material and degradation of polymer), cross-linking (formation of free radicals and branching of macromolecules) and functionalization (formation of new chemical functions) [5–10]. Furthermore, these plasma treatments can be performed at atmospheric pressure, thus avoiding the costs of a vacuum installation and favouring their potential implementation into industrial applications. In particular, He or Ar post-discharges generated by a RF plasma torch demonstrated their efficiency in polymer treatment [8, 11] and in polymer plasma deposition [12]. Duluard et al [13] studied the optical emission of the species produced in an Ar post-discharge enriched in water vapour. They explained the increase in ozone and in nitrogen intensity (second positive system) as resulting from an increase of the torch-to-substrate distance (gap). The plasma activation could be accompanied by a slight etching process (without degradation of the underlying substrate) improving the surface roughness and tending to increase the hydrophilicity and furthermore the adhesion of a future layer.

From the literature, it is commonly admitted that the hydrophilic character enhancement of a polyethylene surface is due to its oxidation which can be achieved by a plasma treatment to produce new oxygen-based functionalities such as hydroxyl or carboxylic groups on the surface [5, 6, 14]. However, this plasma-treated surface is subject to an ageing effect which occurs immediately after the plasma treatment and is considered as finished once the water contact angle (WCA) values reach a plateau. This plateau is usually located from 5° to 25° higher than the values of freshly treated surfaces, depending on the nature of the film: polyethylene terephthalate, polyamide or LDPE [15–17]. This hydrophobic state recovery [18] can be explained by a reorientation of induced polar chemical groups which occurs into the material bulk to minimize the interfacial free energy between the polymer surface and its environment [19]. Moreover, the majority of the free functions or radicals can react either with the oxygen from the ambient air or among themselves. Besides the storage conditions, the ageing of plasma-treated polymer surfaces is also influenced by many other parameters such as the nature of the working gas and the crystallinity of the treated material [20–22].

The present study is focused on the etching/functionalization processes and on the ageing behaviour of plasma-treated LDPE surfaces. For this purpose, an atmospheric-pressure argon plasma torch was used with or without oxygen added as reactive gas. In addition, the influence of the gap and of the oxygen flow rate is evaluated.

## 2. Experimental details

In this study, transparent LDPE films (2×2 cm$^2$) provided by PackOplast-Belgium were used, with a thickness of 40µm and a density of 0.93 g cm$^{-3}$. The polymer was treated using an AtomfloTM-250D plasma source provided by SurfX Technologies LLC [12]. The plasma was generated between an RF powered upper electrode (27.12 MHz) and a lower grounded electrode. The argon flow rate was maintained at 30 L.min$^{-1}$ for all experiments, whereas the oxygen flow rate ($O_2$) was varied between 0 and 25 mL.min$^{-1}$ (at fixed gap). The emissive region of the post-discharge can be assimilated to a plume presenting a length estimated to 1 mm. Due to possible thermal damaging effects, the distance between the LDPE surface and the plasma torch (gap) was tuned between 2 and 30 mm, so that the exposed surfaces were not interacting with the plume. For the ageing study, samples were kept in Petri dishes in ambient conditions (room temperature at 20°C with 72% of humidity).

The dynamic WCA measurements were performed using a Krüss DSA 100 (Drop Shape Analyser) in an air-conditioned room and the working liquid was milli-Q water. Advancing (aWCA) and receding (rWCA) angles were measured by depositing and withdrawing a droplet of 5µl on the surface. In this paper, each value of dynamic WCA corresponds to the average measurements of five drops, randomly deposited onto the sample surface.

The x-ray photoelectron spectroscopy (XPS) measurements were performed with a PHI 5600 photoelectron spectrometer, operating at 300W with a Mg Kα, 1,2 x-ray source (1253.6 eV) and under a vacuum of 9×10$^{-9}$ Torr. Pass energies of the survey spectra and high-resolution spectra were fixed at 93.90 eV and 23.5 eV, respectively. The take-off angle of the photoelectrons was 45° with respect to the sample normal axis. The C1s core level at 285.0 eV was used to calibrate the binding energy scale. The surface elemental composition was calculated after removal of a Shirley background by using the following sensitivity coefficients: $S_C$ = 1 and $S_O$ = 2.85. The peak fitting of the C1s components was carried out with the Casa XPS software.

The mass losses of the plasma-treated LDPE films were evaluated by employing a Sartorius BA110S Basic series analytical balance, characterized by a 110 g capacity and 0.01 mg readability.

The surface roughness was evaluated on images obtained from the atomic force microscopy (AFM) technique. The device was a Dimension 3100 AFM using a Nanoscope IIIa controller equipped with a phase imaging extender, from Digital Instruments operating in the tapping mode (TM-AFM). Standard silicon tips (Tap300Al, BudgetSensors) were used with a 42N.m$^{-1}$ nominal spring constant and a 300 kHz nominal resonance frequency. All images were recorded in air at room temperature with a scan speed of 1 Hz. Except a second-order polynomial function background slope correction, no further filtering was performed. From these flattened corrected data, the root-mean-squared roughness ($R_{rms}$) and the maximum topographic height were determined on the flattened 5 × 5μm$^2$ images.

Optical emission spectroscopy (OES) was performed with a Shamrock 500i from ANDOR Technology equipped with a CCD camera (iDus 420 series camera with 1024 × 255 pixels). A 1800 grooves.mm$^{-1}$ grating blazed at 300 nm (linear dispersion of 1.06 nm.mm$^{-1}$ at a wavelength of 575 nm) was used for optical emission spectra in the visible–near infrared range. The spectra were acquired with an exposure time fixed to 2 ms and a number of accumulations as high as 30.

## 3. Results

### 3.1. Influence of the gap

**3.1.1. Wetting/functionalization**

The variation of the aWCA as a function of the gap is shown in figure 1 in the case of a polyethylene film exposed during 30 s to Ar and Ar-O$_2$ post-discharges powered at 60W. Whatever the nature of the post-discharge, the highest wettability was obtained for the samples treated with the smallest gap. In the case of a pure argon plasma treatment (figure 1(a)), the aWCA decreases from 94° (untreated surface) to 57°±3° for a gap of 10mm and can even reach a value of 43°± 2.5° for a gap of 2 mm. The LDPE films treated with an Ar-O$_2$ post-discharge (figure 1(b)) depicts the same behaviour with values as low as 41° for a gap of 2 mm. For both plasma treatments, one can emphasize that the LDPE surface becomes more hydrophilic for a gap lower than 15 mm.

In order to verify whether the surface activation could be achieved by polar groups (i.e. incorporation of oxygen functions responsible for the increasing hydrophilicity) [6, 14, 23, 24], the atomic oxygen concentration was determined from the area estimated under the O1s peak from the wide XPS spectra. In figure 1(a) as in figure 1(b), a small amount of oxygen is detected on the films treated for gaps higher than 15mm while more oxygen is implanted on smaller gaps. Indeed, at 2 mm, the oxygen surface concentration reaches almost 25.0% and 29.5% in Ar and Ar-O$_2$ post-discharges, respectively. This small change in the oxygen surface concentration between these two treatments could be explained by the etching process occurring on the topmost surface layers. A correlation between the etching rate and the oxygen flow rate is suggested later in this paper.

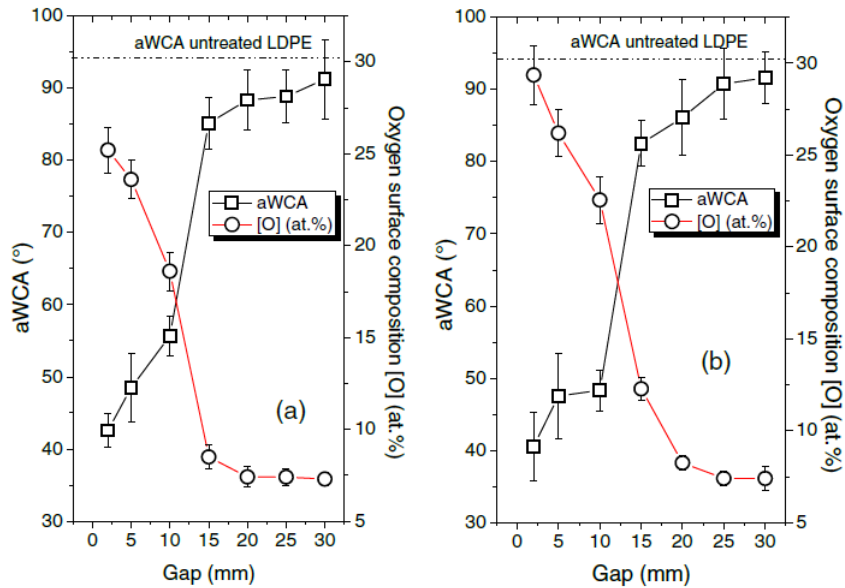

*Figure 1. Advancing WCAs and atomic oxygen concentration of LDPE films treated for (a) a pure Ar and (b) an Ar-$O_2$ post-discharge (t = 30 s, P = 60W, $\Phi_{O2}$ = 15 mL.min$^{-1}$ (b)).*

### 3.1.2. Topography/etching

In order to investigate topographical changes of the plasma-treated surfaces, AFM imaging was achieved in tapping mode and compared to mass losses measurements performed on the samples after exposure to the post-discharge. The $R_{rms}$ values and the relative mass losses for Ar and Ar-$O_2$ post-discharge treatments as a function of the gap are shown in figures 2(a) and (b), respectively.

In figure 2(a), for a gap ranging between 2 and 30 mm, a decrease in the $R_{rms}$ is observed from 42.5 to 1.8 nm, this last value corresponding to the pristine state. For gaps lower than 15 mm, the high $R_{rms}$ values suggest a significant etching of the surface induced by the plasma treatment. Figure 2(a) also indicates a mass loss close to 35 µg.cm$^{-2}$ for a gap of 2mm while almost no material was removed for gaps higher than 15 mm. A similar behaviour can be observed in the case of an Ar-$O_2$ plasma treatment where the mass loss turns around 49 µg.cm$^{-2}$ for a gap of 2mm (figure 2(b)). The correlation between the surface roughness values and the mass loss measurements could indicate an enhancement of the surface etching induced by a decrease of the gap.

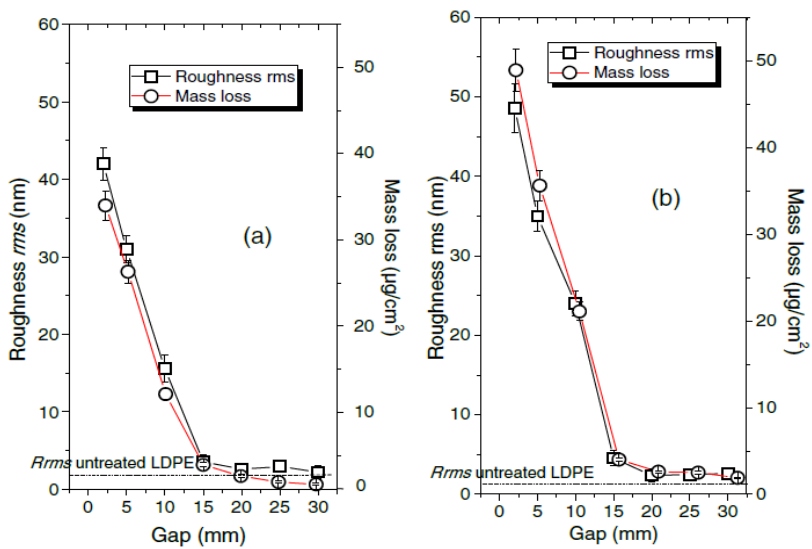

*Figure 2. $R_{rms}$ values and mass losses of plasma-treated films (t = 30 s, P = 60W, $\Phi_{O2}$ = 15 mL.min$^{-1}$) in the case of (a) pure Ar and (b) Ar-$O_2$ post-discharges.*

### 3.1.3. The plasma phase

OES was carried out in order to determine the species from the plasma phase which could contribute to the surface modification. For a gap of 10 mm, no ion could be detected in the Ar or in the Ar-$O_2$ post-discharge, but only singlet Ar, O, N, OH and NO species. The intensities of O (777.7 nm), Ar (772.9 nm), $N_2$ (380.8 nm) and NO (274.2 nm) are plotted as a function of the gap in figures 3(a) and (b). We also reported in figure 4 the emission spectra of $N_2$ (C–B, 0–2) and NO ($B^2\Sigma - X^2\Pi$) respectively detected at 380.8 nm and 274.2 nm. In figure 3(a) as in figure 3(b), the intensities of the Ar and O species decrease monotonically as a function of the gap, which is consistent with the assumption of their production within the discharge. The intensities of these emitted species are globally lower in the Ar post-discharge than in the Ar-$O_2$ post-discharge. The intensities of the $N_2$ and NO bands increase between 2 and 10mm for the Ar and the Ar-$O_2$ post-discharges, but then continuously decrease beyond these two thresholds. Duluard et al—who worked on the same RF plasma torch setup—suggested that theN2 molecules diffusing into the post-discharge were either completely dissociated or flowed with the injected carrier gas [13]. As Duluard et al detected a very low intensity of the $N_2$(C) species for a gap lower than 7 mm, the air diffusion was assumed to be weak in the range of 2–5 mm, due to the elevated argon flow rate (30 L.min$^{-1}$).

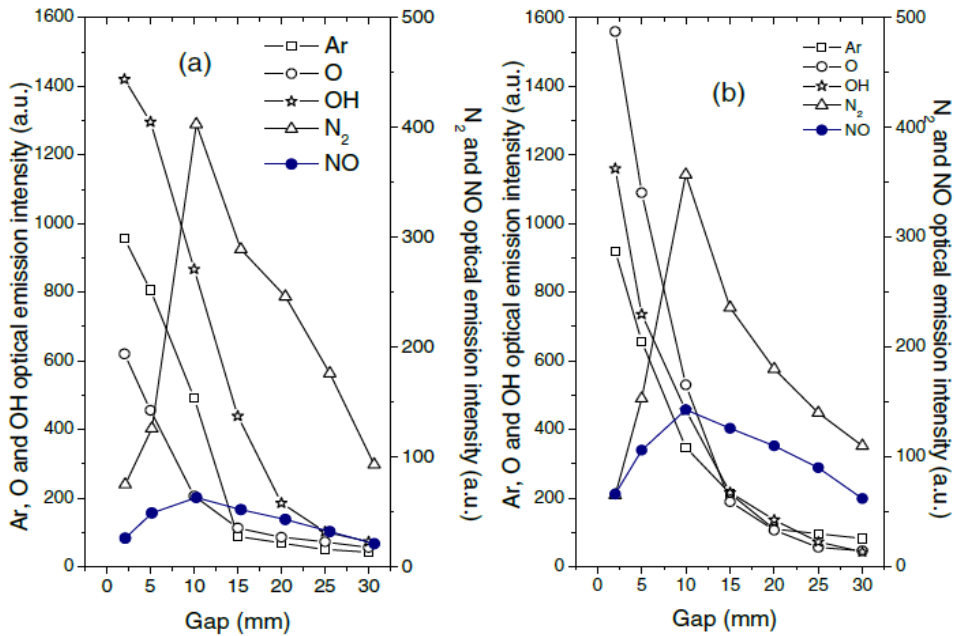

Figure 3. Peak intensities of the species observed in the post-discharge (a) versus the gap in pure argon plasma at 60W, (b) versus the gap in an Ar-$O_2$ plasma with $\Phi_{O2}$ = 15 mL.min$^{-1}$ at 60W.

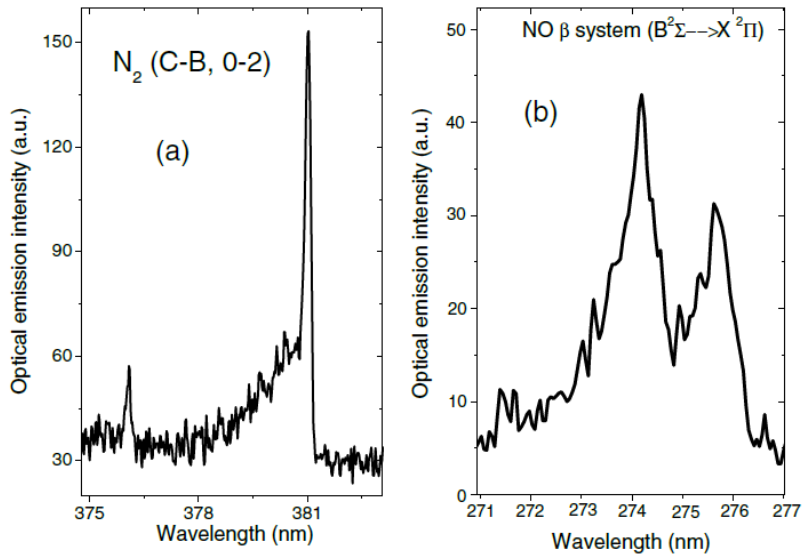

Figure 4. Emissive bands of $N_2$ and NO measured in the Ar-$O_2$ flowing post-discharge for $\Phi_{Ar}$ = 30 L.min$^{-1}$, $\Phi_{O_2}$ = 15 mL.min$^{-1}$, P = 60W, gap = 5 mm.

## 3.2. Influence of the reactive gas flow ($O_2$)

In this section, the hydrophilicity, the roughness and the etching mechanisms of the plasma-treated LDPE surfaces were studied as a function of the oxygen flow rate. The gap was fixed to 10mm since beyond this value the plasma–surface interactions were too weak, and below (lower than 5 mm), the treated films sustained significant degradations attributable to thermal effects.

**3.2.1. Wetting/functionalization**

Figure 5 reports the variations of the dynamic WCA but also the oxygen and nitrogen relative surface concentrations versus the oxygen flow rate, for post-discharge treatments with a gap fixed at 10 mm, an RF power of 60W and a treatment time set to 30 s. For an increase in the oxygen flow rate from 0 to 5 mL.min$^{-1}$, the aWCA decreases while for higher flow rates, a plateau close to 47° is reached. In contrast, the receding WCA (rWCA) continuously decreases over the 0–25 mL.min$^{-1}$ range. For comparison, the XPS spectrum of the untreated LDPE is only dominated by the C1s peak. The wide survey of the treated LDPE shows the presence of both C1s and O1s peaks, and to a lesser extent nitrogen N 1s. Besides, the increase in the $O_2$ flow rate from 0 to 25 mL.min$^{-1}$ demonstrates a higher incorporation of oxygen from 21% to 26% and a small increase in the nitrogen concentration from 0.5% (no oxygen) to 2.5% (25 mL.min$^{-1}$ of oxygen). The surface of the films freshly treated by the Ar-$O_2$ post-discharge tends to be saturated in oxygen for $\Phi_{O_2}$ > 15 mL.min$^{-1}$.

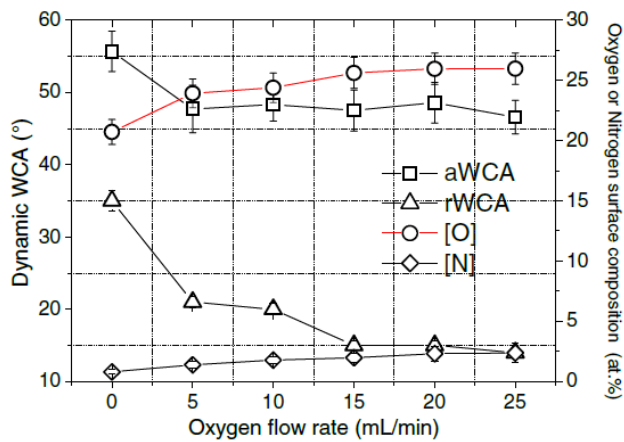

Figure 5. Dynamic WCA (a = advancing and r = receding), atomic oxygen and nitrogen concentrations of plasma-treated films (t = 30 s, P = 60W, gap = 10 mm) in terms of the oxygen flow rate.

### 3.2.2. Topography/etching

Figure 6 illustrates AFM images of (a) an untreated LDPE surface and (b), (c) several plasma treated LDPE surfaces for 5, 15 and 25 mL.min$^{-1}$ in the O$_2$ flow rate. The corresponding $R_{rms}$ values of these pictures, but also of the pictures obtained at 10 and 20 mL.min$^{-1}$ of O$_2$, are reported as a function of the oxygen flow rate in figure 7. Whatever the O2 flow rate, the $R_{rms}$ values are always much higher than the $R_{rms}$ of a native surface, which is only 1.8 ± 0.2 nm. In a pure Ar post-discharge, the $R_{rms}$ is 5.8±0.3 nm and increases until 36 nm for $\Phi_{O2}$ = 25 mL.min$^{-1}$. Furthermore, the mass loss of the plasma-treated surface reported in figure 7 follows a linear increase with the O$_2$ flow rate. A close correlation can therefore be suggested between the roughness and the mass loss of the sample.

Figure 8 shows typical mass losses of the sample after its exposure to the post-discharge as a function of the treatment time for several oxygen flow rates. As the mass loss varies linearly with the plasma exposure time, a mean etching rate (nm.s$^{-1}$) was evaluated by calculating the slope of each linear fit. Knowing the density of the LDPE (0.93 g.cm$^{-3}$), the mass loss (expressed in µg.cm$^{-2}$) could easily be converted to a thickness (nm). The etching rates are plotted versus the oxygen flow rate in figure 8(b). The mean etching rate is linearly increasing with the oxygen flow rate, with a maximum as high as 7.3 nm.s$^{-1}$ reached for $\Phi_{O2}$ = 25 mL.min$^{-1}$.

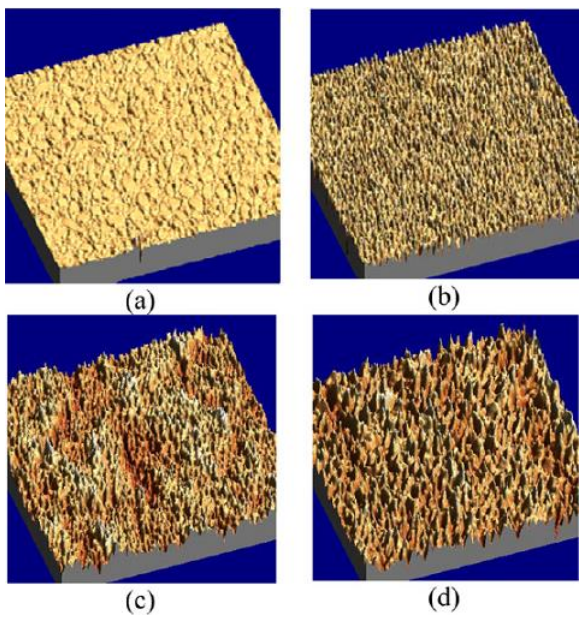

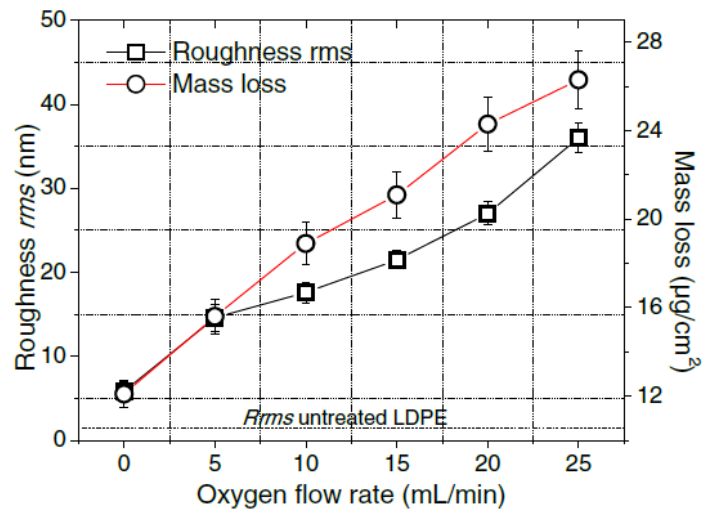

Figure 6. 3D AFM images (5 × 5µm$^2$) of (a) untreated and treated (t = 30 s, P = 60W, gap = 10 mm) LDPE for different flow rates: (b) 5 mL.min$^{-1}$, (c) 15 mL.min$^{-1}$ and (d) 25 mL.min$^{-1}$.

Figure 7. $R_{rms}$ values and the mass loss observed on the surface of the treated films (t = 30 s, P = 60W, gap = 10 mm) in terms of the oxygen flow rate.

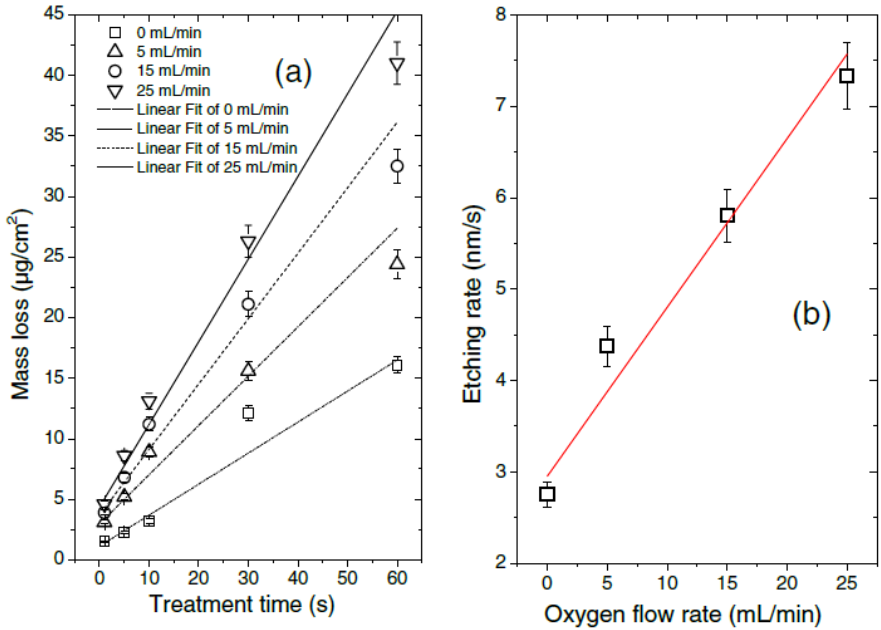

Figure 8. (a) Mass loss of the LDPE films after exposure to an Ar-O$_2$ post-discharge ($\Phi_{O2}$ = 0–25 mL.min$^{-1}$) as a function of the treatment time. (b) Mean etching rate of the LDPE films as a function of oxygen flow rate. The films were treated with a gap of 10 mm.

### 3.2.3. The plasma phase

The intensities of O (777.7 nm), Ar (772.9 nm), N$_2$ (380.8 nm) and NO (274.2 nm) are plotted as a function of the oxygen flow rate in figure 9. The production of O and NO species (increasing curves) is balanced by the consumption of Ar, OH and N$_2$ species (decreasing curves). The possible reactions are reported in table 2 and detailed in the discussion section.

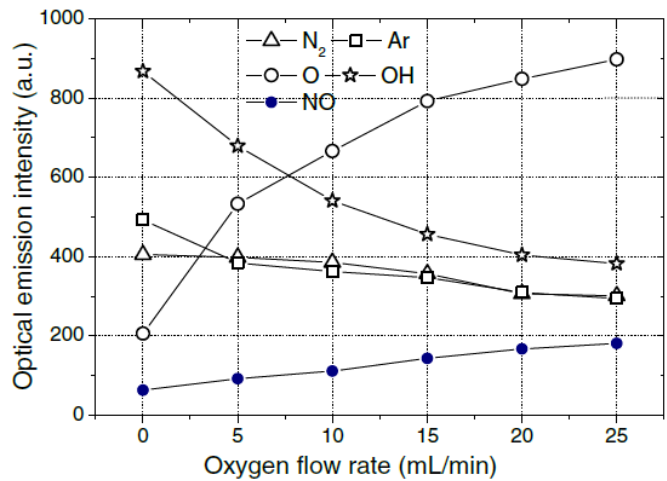

Figure 9. Peak intensities of the species observed in the post-discharge versus the oxygen flow rate for a fixed gap of 10mm at 60W.

### 3.3. Stability of the modified surfaces

As the atmospheric plasma treatment considerably enhanced the amount of polar groups at the polymer surface, a dedicated study was achieved to investigate the evolution of the surface wettability over time. The plasma-treated films were subjected to the air ageing (72% RH, 20°C) to evaluate their hydrophobic recovery time, as shown in figure 10(a), where the aWCA are plotted as a function of the oxygen flowrate. Immediately after the post-discharge treatment (gap of 10mm and RF power of 60 W), the LDPE film freshly treated by an Ar post-discharge becomes less hydrophilic (55.7±1.7°) than a film treated by an Ar-$O_2$ post-discharge (47.5 ± 2.8°). This difference between Ar and Ar-$O_2$ post-discharge treatments vanishes with ageing time. In figure 10(b), whatever the oxygen flow rate, the ageing process can be regarded as completed after 30 days since the aWCA values reach a plateau close to 80°. Even after 60 days of ageing, the effect of the plasma treatment is still permanent since the treated LDPE surfaces remain more hydrophilic (about 80°) than the untreated surfaces (close to 94°).

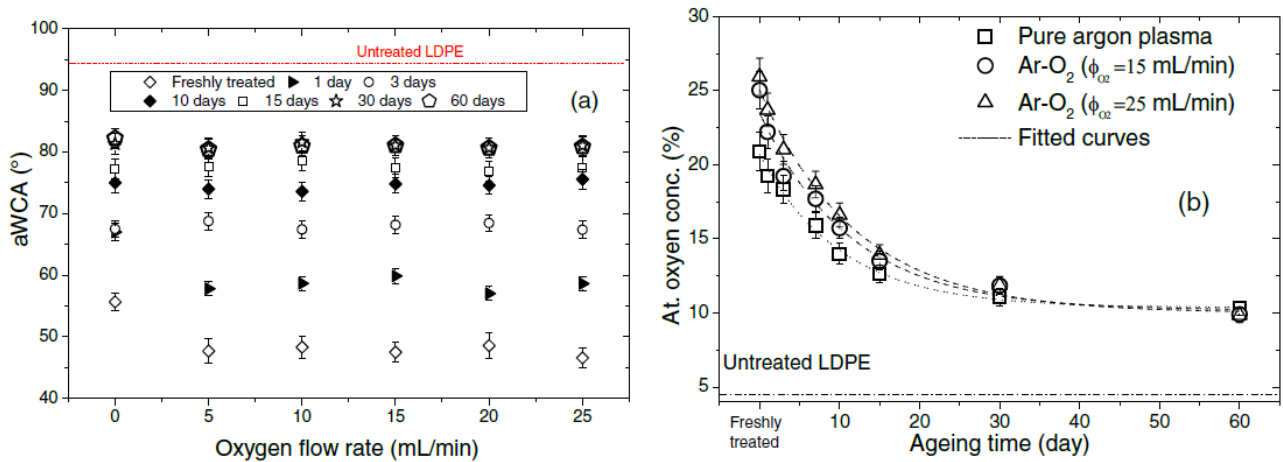

Figure 10. (a) aWCA measurements of plasma-treated films versus the ageing time, (b) atomic oxygen concentration for films treated by an Ar-$O_2$ post-discharge with $O_2$ equal to 0, 15 and 25 mL.min$^{-1}$.

These aWCA measurements can be correlated with the chemical surface composition detected by XPS. Figure 10(b) shows the relative elemental concentration of atomic oxygen present on the LDPE surface as a function of the ageing time. Here again, the films treated by the post-discharge with the highest $O_2$ flow rate reach the final hydrophilic state faster. Whatever the initial amount of oxygen measured on the surface, they all reach a concentration close to 12% after 30 days.

It is noticeable that the variation of the oxygen concentration is in an exponential decay:

$$[O] = k.\exp(-t/\tau) + [O]_0,$$

where $[O]_0$ is the offset corresponding to $[O]$ when t tends to the infinity, k is the amplitude and τ is the decay constant (or time constant) in terms of the ageing time. The variation of the oxygen concentration can be fitted with a first-order linear time invariant. From the fitted curves (figure 10(b)), the constant time (τ) was evaluated to 10.17, 11.42 and 12.04 days for pure argon plasma, Ar-$O_2$ plasma ($\Phi_{O_2}$ = 15 mL.min$^{-1}$) and Ar-$O_2$ plasma ($\Phi_{O_2}$ = 25 mL.min$^{-1}$), respectively. '$[O]_0$' (representing the plateau reached in figure 10(b)) presents the same value (12%) whatever the post-discharge treatment, since the LDPE-treated surface keeps the same oxygen concentration after the ageing process, whatever the plasmas conditions. The correlation between the XPS results and the aWCA measurements demonstrate that the oxygen functionalities are mainly responsible for the hydrophilic state of the plasma-treated surfaces.

To evaluate the chemical composition modifications of the plasma-treated LDPE surfaces, figure 11 introduces two XPS survey spectra: (a) is for the untreated surface and (b) for the surface treated by an Ar post-discharge. High-resolution peak fitting on C1s peaks are also plotted in both cases to highlight the nature of the chemical bonds supported by the C atoms. Before the plasma treatment, the C1s peak contains three components corresponding to (i) the hydrocarbon bond (285.0 eV), (ii) the carbon oxide bond (286.5 eV) and (iii) the carbonyl bond (288.0 eV). After the plasma treatment, the C1s peak contains a fourth contribution, namely a carboxyl carbon bond (288.9 eV) [21].

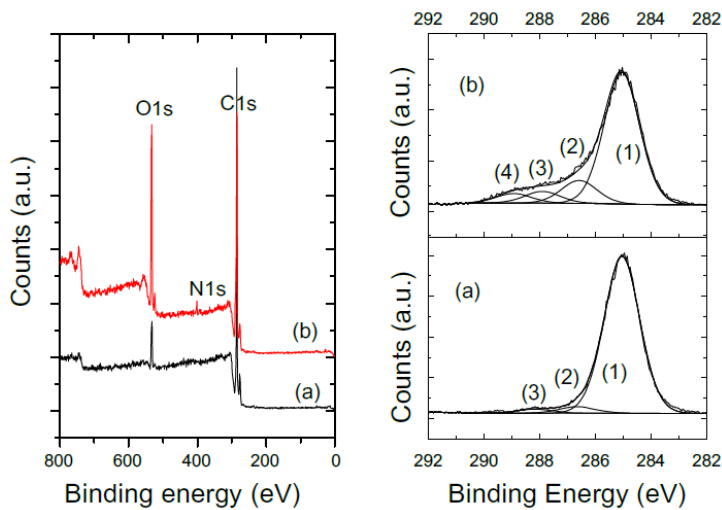

*Figure 11. XPS survey spectra for (a) an untreated LDPE and for (b) a LDPE surface treated with an Ar post-discharge (t = 30 s, P = 60W, and gap = 10 mm, P = 60 W). High-resolution C1s peaks are decomposed for each case.*

The presence of nitrogen (<3%) on the plasma-treated surface may slightly influence the oxygenated contributions defined at 286.5 eV and 288.0 eV as these bond energies are also specific to (C–N, C=N, C≡N) and (N–CO–N, N–C–O, N–C=O) groups, respectively [25]. Since the energy difference between some nitrogen- and oxygen-related groups is too small to allow a mathematical separation within error bars, the identification of the functional carbons based on the binding energies of the N1s and O1s peaks is rather difficult. Besides, the N1s peak lies between 397 and 403 eV: an energy range where several carbon–nitrogen species such as amides, imides, nitriles can be found. The peak fitting of N1s cannot therefore accurately specify the nature of the nitrogen containing functionalities on the LDPE surface. According to table 1, the nitrogen surface concentration is vanishing with the ageing time for a post-discharge treatment of 30 s and an $O_2$ flow rate of 15 mL.min$^{-1}$. The same day after the plasma exposure, the nitrogen content is 2.8% while it is lower than 0.5% 10 days afterwards.

| Ageing time (day) | [N] (at%) |
|---|---|
| 0 | 2.8 |
| 1 | 2.1 |
| 3 | 1.2 |
| 10 | <0.5 |
| 15 | <0.5 |
| 30 | <0.5 |
| 60 | <0.5 |

*Table 1. Atomic nitrogen concentration taken from XPS analysis of an aged LDPE film, treated in the following post-discharge conditions: RF power = 60W, treatment time = 30 s, $\Phi_{O2}$ = 15 mL.min$^{-1}$.*

Figure 12 compares the ageing time of the C–C (and/or C–H), C–O, C=O and O–C=O bond concentrations measured on a treated LDPE surface either by (a) an Ar or (b) an Ar–$O_2$ post-discharge. Whatever the treatment, it appears that the relative proportion of the oxygenated functions decrease with the ageing time, to the benefit of the C–C and/or C–H bonds.

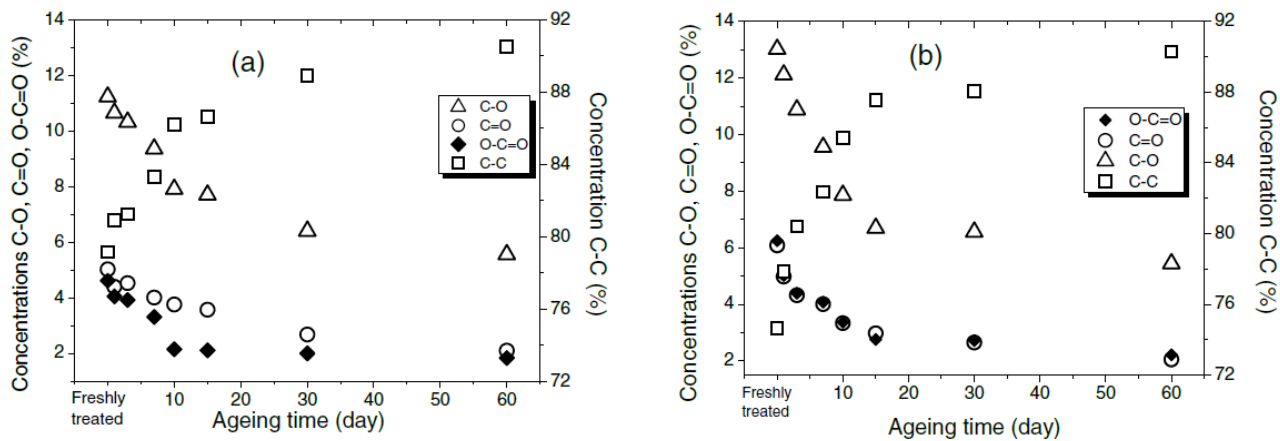

*Figure 12. Concentrations of the different carbon bonds versus the ageing time for samples treated with (a) a pure argon post-discharge, (b) with an Ar-O$_2$ post-discharge (t = 30 s, P = 60W, $\Phi_{O_2}$ = 15 mL.min$^{-1}$, gap = 10 mm).*

# 4. Discussion

## 4.1. The plasma phase

Surfaces treated by an atmospheric plasma torch sustain milder conditions than in a low-pressure plasma source since the lifetime of the reactive species is shorter at atmospheric pressure and also due to the UV radiation which is less emissive in a post-discharge. As a consequence, the species produced in the discharge lose their reactivity upon an increase in the gap and only the long-life species can diffuse and reach the polymer surface to participate in the incorporation of functional groups [15, 26, 27].

Several mechanisms are usually considered to explain the production of O species, even if all cannot apply here. At atmospheric pressure, the argon metastable species are mostly produced within the discharge by collisions between electrons and argon atoms, according to reaction (1) in table 2. The argon metastable species from the discharge could be consumed by quenching with O$_2$ or O species according to reactions (2) and (3) respectively, or by collision with two argon atoms to produce molecular argon (Ar2) according to reaction (4). The production of O atoms can be achieved either by the electronic dissociation of O$_2$ or assisted by the argon metastable species as suggested in reaction (2). As no band of O$_2$ excited molecules was detected, not even O$_2$ metastable species between 750 and 770 nm [40], a complete dissociation of O$_2$ in O radicals can be assumed and reaction (5) seems a good candidate to explain the possible production of O$_3$. The O (1D) oxygen species may de-excite towards their fundamental level (3P) through the collisional process (6). They cannot be involved in the destruction of O$_3$ according to reaction (7) since no emission of O$_2$ metastables was detected between 750 and 770 nm. The radiative decay of N$_2$ reported in reaction (8) is validated since the transition (B 3 → A 3 ) was evidenced by OES at 654 nm. Furthermore, the N$_2$ molecules can recombine with O$_2$ to produce NO molecules as suggested in reaction (9) and as observed experimentally in figure 3. No N atoms were detected by OES either because no N atom was produced or because some of them are produced but consumed upon a reaction presenting a very elevated rate coefficient. The hypothetical excess in N atoms could be consumed by O$_2$ to produce NO and O species (reaction (10)). This observation is consistent with the increase in intensity of NO and O atoms when the gap is increased.

| | Reaction | Name | Rate coefficient ($cm^3 s^{-1}$) |
|---|---|---|---|
| (1) | $e^- + Ar \rightarrow e^- + Ar_m^*$ | Excitation to $Ar_m^*$ | — |
| (2) | $Ar_m^* + O_2 \rightarrow Ar + O + O$ | Quenching by dissociation of $O_2$ | $2.1 \times 10^{-10}$ |
| (3) | $Ar_m^* + O \rightarrow Ar + O$ | Quenching of O | $4.1 \times 10^{-11}$ |
| (4) | $Ar_m^* + 2\,Ar \rightarrow Ar_2 + Ar$ | Three-body collisions | $1.4 \times 10^{-32}$ |
| (5) | $O\,(^3P) + O_2(X) + Ar \rightarrow O_3 + Ar$ | Ozone formation | $1.9 \times 10^{-35}\,e^{(1057/T_g)}\,cm^6.s^{-1}$ |
| (6) | $O(^1D) + O(^3P) \rightarrow 2O(^3P)$ | Collisional de-excitation | $8 \times 10^{-12}$ |
| (7) | $O\,(^1D) + O_3 \rightarrow 2O\,(^3P) + O_2(X)$ | Ozone destruction | $1 \times 10^{-12}$ |
| (8) | $N_2(B\,^3\Pi_g) \rightarrow N_2(A\,^3\Sigma_u^+) + h\nu$ | Radiative decay | $2.W43 \times 10^{-5}$ |
| (9) | $N_2(B\,^3\Pi_g) + O_2 \rightarrow 2NO$ | Recombination | $1.2 \times 10^{-10}$ |
| (10) | $N(^4S) + O_2 \rightarrow NO + O(^3P)$ | Oxygen and nitrogen destruction | $8.9 \times 10^{-17}$ |

Table 2. Partial list of the chemical reactions with their rate coefficients for the Ar-$O_2$ flowing post-discharge.

At atmospheric pressure, OH radicals are produced mainly (i) by electron-impact dissociation of H2O molecules or (ii) by electron-impact ionization of $H_2O$ followed by dissociation of $H_2O^+$ to produce OH. As these reactions require energetic electrons they probably do not occur within the post-discharge. The production of OH radicals is assumed to be performed between the electrodes. The injection of $O_2$ in the post-discharge is assumed to decrease the electron density, thus limiting the production of OH radicals and therefore their emission intensity. For this reason, a decrease in the OH emission is observed in figure 9 as a function of the increasing $O_2$ flow rate.

In a non-thermal plasma, two types of active species can be distinguished: the ones which are chemically reactive (for example O* and N*) and the others which are non-reactive but able to break chemical bonds (photons, electrons, non-reactive ions and non-reactive excited species) provided a sufficient energy [46, 47]. The comparison between the energy of the species produced within the post-discharge (Ar, NO, O) and the bond dissociation energies (in the LDPE structure) reported in table 3, suggests that the O radicals, excited states of Ar atoms and $N_2^+$ ions have enough energy to be responsible for the etching process occurring during the treatment. The Arm species could be sufficiently energetic but could not be detected in the post-discharge.

| LDPE polymer | Bond energy (eV) | Ref. | Species from the post-discharge | Upper energy level (eV) |
|---|---|---|---|---|
| C–H | 4.81 | [41] | O | 10.7 |
| C–C | 3.65 | [41] | $Ar^m$ | 11.6 |
| | | | Ar I (706.7 nm) | 13.3 |
| | | | Ar I (763.5 nm) | 13.2 |
| | | | Ar I (811.5 nm) | 13.1 |
| | | | $N_2^+$ | 18.6 |
| | | | NO | 4.5 |
| | | | $O_2(b\,^1\Sigma_g^+)$ | 1.63 |
| | | | $O_2(a\,^1\Delta_g)$ | 0.98 |

Table 3. Bond dissociation energies in the LDPE and upper energy levels of the main species present in the Ar-$O_2$ post-discharge.

## 4.2. Competitive and synergetic surface processes: activation, etching and oxygen diffusion

During the plasma exposure, two simultaneous competitive processes occurred: the etching (figure 8) and the activation (figure 5) of the surface. The diffusion of oxygen could also participate as an additional process during the plasma treatment [48] provided a prior surface saturation in oxygenated functional groups.

**4.2.1. Etching process**

As expected, the LDPE surface becomes much rougher after Ar or Ar-$O_2$ plasma treatments. The increase in the $R_{rms}$ of the plasma-treated films correlated with the mass loss measurements indicated a significant etching process of the films. Similar linear behaviours have already been observed in other polymers [11]. For instance, in the case of a polytetrafluoroethylene (PTFE) sample treated by an atmospheric He-$O_2$ plasma torch, Hubert et al correlated a variation $\Delta R_{rms}$ of 45 nm with a mass loss variation $\Delta m$ of −550 ppm [49]. The mass losses were greater if oxygen was mixed with the carrier gas.

In our case, the mean etching rate, estimated between 2.7 and 7.3 $nm.s^{-1}$, is much higher than those observed for other polymers. Indeed Inagaki et al [4] who treated PET surfaces with a density of 1.38 $g.cm^{-3}$, found a lower etching rate (close to 1.15 $nm\ s^{-1}$) that may be attributed to their experimental conditions: an RF low-pressure argon plasma (13.3 Pa). Moreover, the density of the treated polymer was higher than the density of the LDPE studied here. For the sake of comparison, table 4 reports various etching rates corresponding to different plasma treatments.

The various etching rates from table 4 seem to indicate that they do not strongly depend on the pressure since they all are on the order of few $nm.s^{-1}$ for pressures ranging between 13 Pa and atmospheric pressure. The detected species with a high energy (such as $O_I$: 10.74 eV) can be responsible for a physical etching since the bond dissociation energies of C–H and C–C are 4.81 eV and 3.65 eV, respectively.

The roughness formation can be due to two randomly distributed phases with different etch rates. Two etching modes could be defined/examined as follows: (1) with anisotropic flux of species (sputtering) and (2) with isotropic flux of etchants (chemical etching), as explained in the works of Zakka et al [52]. The roughness in our case is the anisotropic etching mode. The roughness formation mechanism could rely on a natural self-organization mechanism occurring during the erosion of surfaces, based on the interplay between roughening induced by energetic species etching and cross-linking due to surface reorganization [53]. It could also result from the existence of local crystalline micro-areas on the surface which require more energetic species from the post-discharge than the amorphous regions to be etched. This assumption has been supported by Hubert et al to explain the texturization of PTFE surfaces induced by O radicals produced from a He-$O_2$ flowing post-discharge at atmospheric pressure [11].

In figure 5, the hysteresis resulting from the aWCA and the rWCA versus the oxygen flow rate is in agreement with a Wenzel state, where the water remains trapped in the asperities resulting from the etching process [54]. As reported in table 5, the hysteresis can be correlated with the surface roughness but also with surface composition heterogeneities (particularly when a polymer surface contains both hydrophobic and hydrophilic groups) and surface rearrangements during the contact with water. As a result, two parallel synergetic phenomena occur on the surface: the incorporation of oxygen functional groups promoting its hydrophilic character and the etching process promoting its roughening. Bico et al studied the influence of the roughness on the surface hydrophilicity [55] and showed that a rougher hydrophilic surface would become more hydrophilic than a smooth surface.

#### 4.2.2. Activation

In the literature [14, 17], different treatment times—from few milliseconds to several minutes—are reported for polymer surface modifications. At the beginning of the treatment, contaminants from the polymer surface can be removed, which may also lead to improved wettability. With further treatment time, the insertion of oxygen/nitrogen atoms on active sites of the polymer surface can occur, leading to the formation of various functional groups, thus changing the surface wettability. For longer treatment times, excessive scissions may take place leading to a layer of low molecular weight fragments on the surface such as LMWOM (low molecular weight oxidized material) as discussed afterwards.

In our case, the plasma activation resulted in the insertion of active species (oxygen containing species) from the radicals of the post-discharge. This phenomenon was more pronounced on samples exposed to a post-discharge enriched in oxygen (oxygen flow rate >5 mL.min$^{-1}$) since greater amounts of oxygen radicals were generated. The O and OH radicals from the post-discharge are known to abstract secondary hydrogen atoms from LDPE polymer chains, resulting in the formation of alkyl radicals, as indicated in reaction (11). These alkyl radicals can react with atomic oxygen or ozone and lead to the formation of alkoxyl radicals. As Duluard et al [13] showed, the increase of the ozone concentration by increasing the oxygen flow rate with the same RF plasma torch, reaction (11) could explain the increasing incorporation of oxygen functions onto the LDPE surface.

Moreover, C-radical sites from LDPE could react with oxygen radicals from the ambient air and result in the formation of peroxy radicals (reaction (12)) and hydroperoxides (reaction (13)) [21]. During the plasma treatment, chains scissions occur while chains with radical tails having more mobility are created. The breaking of some polymer chains can induce surface rearrangements such as polymer crystallinity modifications [23–56]. Immediately after post-discharge exposure, radicals from the surface can react with the oxygen from the ambient atmosphere and enhance the surface activation [57].

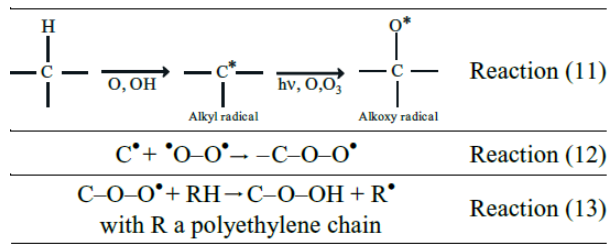

### 4.3. Ageing study

In figure 12, the relative proportion of the oxygenated functions decreases with the ageing time to the benefit of the C–C and/or C–H functions. This result correlated with the increase in the aWCA obtained during the ageing process (figure 10(a)) may be attributed to a reorientation of the induced polar groups into the bulk of the polymer, whatever the plasma treatment. The treated surface would reorganize itself immediately after the Ar or Ar–O2 plasma treatment, without—nevertheless—recovering its native hydrophilic state. Besides, as reported in table 1, the nitrogen species could be ejected from the surface after the treatment under the form of NOx gases, since no nitrogen was detected by XPS on the surfaces of the films aged more than 10 days.

In figure 10(a), the sharp increase observed in aWCA in one day was similar with the sharp decay of atomic oxygen concentration from the XPS curve in figure 10(b). According to Gerenser et al, low molecular weight oxidized material (LMWOM) coming from chain scissions during the plasma treatment, could slowly diffuse into the bulk material upon the ageing time, thus contributing to decrease the amount of oxygen on the topmost layers of the surface film [58]. The formation of non-volatile oligomers (i.e. LMWOM) may dominate the etching process, depending on the polymer, the reactive gas and the discharge conditions. The ageing study of the treated surfaces reveals a well-known [20] and partial recovery of the native wettability state, resulting from the rearrangement of polar species and the migration of LMWOM after a few days. The post-discharge also removes LMWOM or converts them into high molecular weight oxidized material (HMWOM) through cross-linking reactions. As a result, the weakly bound layers formed by the LMWOM are removed [59, 60]. This recovery remained partial since even two months later, the highest aWCA turned around 80° instead of 94° (native LDPE). An inert gas discharge can induce a so-called CASING (cross-linking by activated species of inert gases) process, creating a cross-linked layer on the polymer surface [60]. On the one hand, the formation of this layer due to the plasma treatment can subsequently restrict the chain mobility and thus partially inhibit the recovery of the native wettability state. The polar groups, which are likely to reorient into the bulk, are confronted to a cross-linking of the top layer. In the case of the ultra-high modulus polyethylene (UHMPE), the wettability seemed to decrease linearly with the ageing time [61]. On the other hand, the formation of a rougher surface after the exposure to the post-discharge causes an increase in hydrophilicity (lower contact angle), in agreement with the Wenzel model. The roughness of surface area seems to be stable with ageing time and this stability always creates a lower contact angle than the untreated LDPE one. After 40 days of storage, the UHMPE WCA was close to 90°, without nevertheless reaching any plateau that could indicate the end of the ageing process (contrarily to our case). The slow ageing rate (increase of aWCA with the storage time) obtained in this paper, can therefore be considered as a better result. What is more, the induced treatment is favourable to the good adherence [62, 63] between the plasma-treated LDPE film and a subsequent layer to be deposited.

## 5. Conclusion

The correlation between the aWCA, XPS and AFM results showed a better functionalization and roughness in the case of an Ar-$O_2$ post-discharge than for a pure Ar post-discharge. The Ar-$O_2$ post-discharge seems to create the same functional polar groups but at higher concentrations and to promote the surface roughness, hence a better wettability state. The reactive gas flowrate as well as the gap influenced the chemical composition of the LDPE surface, but also its morphology and its wettability. Etching rates (between 2.7 and 7.3 nm.s$^{-1}$) specific to the plasma treatment conditions were determined by correlating mass loss measurements (between 10 and 60 µg.cm$^{-2}$) with treatment times. The improvement of the hydrophilicity was greater for the smaller gaps (lower than 15 mm) where—according to our OES results—the more energetic species were able to break bonds and either eject fragments from the surface or create surface radicals. The ageing study showed that the aWCA of the modified surfaces increased towards a common threshold of 83° after 30 days of storage, whatever the $O_2$ flow rate, thus indicating a partial recovery of the native wettability state. The formation of LMWOM or of HMWOM through the CASING process could explain the fact that the surface was not stable over a period shorter than 30 days. Furthermore, we showed with the ageing time, that the functionalization was partially vanishing while the surface roughness was preserved. As a consequence, the fact that the native wettability state was never totally recovered could be attributed to this persisting roughness but also to the presence of remaining polar groups reoriented into the bulk.

## 6. Acknowledgments

We gratefully acknowledge the support of the Everwall Project financed by the FEDER–Region Wallonne and Europe. We also thank the IAP (Interuniversity Attraction Pole) P7/34 program funded by the Belgian Federal Government.